\documentclass[twocolumn,showpacs,preprintnumbers,amsmath,amssymb,superscriptaddress]{revtex4-1}
\usepackage[utf8]{inputenc}
\usepackage{graphicx}
\usepackage{subfigure, psfrag, amsmath}
\usepackage{verbatim}

\begin{document}

\title{Failure of adaptive self-organized criticality during epileptic seizure attacks}

\author{Christian Meisel}
\affiliation{Biological Physics Section, Max-Planck-Institut f\"{u}r Physik komplexer Systeme, N\"{o}thnitzer Stra\ss e 38, 01187 Dresden, Germany}
\affiliation{Department of Neurology, University Clinic Carl Gustav Carus, Fetscherstra\ss e 74, 01307 Dresden, Germany}
\author{Alexander Storch}
\affiliation{Department of Neurology, University Clinic Carl Gustav Carus, Fetscherstra\ss e 74, 01307 Dresden, Germany}
\author{Susanne Hallmeyer-Elgner}
\affiliation{Department of Neurology, University Clinic Carl Gustav Carus, Fetscherstra\ss e 74, 01307 Dresden, Germany}
\author{Ed Bullmore}
\affiliation{Behavioural \& Clinical Neuroscience Institute, Departments of Experimental Psychology and Psychiatry, University of Cambridge, Cambridge, United Kingdom}
\affiliation{Clinical Unit Cambridge, GlaxoSmithKline, Addenbrooke's Hospital, Cambridge, United Kingdom}
\author{Thilo Gross}
\affiliation{Biological Physics Section, Max-Planck-Institut f\"{u}r Physik komplexer Systeme, N\"{o}thnitzer Stra\ss e 38, 01187 Dresden, Germany}

\date{\today}

\begin{abstract}
Critical dynamics are assumed to be an attractive mode for normal brain functioning as information processing and computational capabilities are found to be optimized there. Recent experimental observations of neuronal activity patterns following power-law distributions, a hallmark of systems at a critical state, have led to the hypothesis that human brain dynamics could be poised at a phase transition between ordered and disordered activity. A so far unresolved question concerns the medical significance of critical brain activity and how it relates to pathological conditions. Using data from invasive electroencephalogram recordings from humans we show that during epileptic seizure attacks neuronal activity patterns deviate from the normally observed power-law distribution characterizing critical dynamics. 
The comparison of these observations to results from a computational model exhibiting self-organized criticality (SOC) based on adaptive networks allows further insights into the underlying dynamics. Together these results suggest that brain dynamics deviates from criticality during seizures caused by the failure of adaptive SOC. 
\end{abstract}

\maketitle

\section{Introduction}

In the terminology of physics, a system is said to be in a critical state if it is poised on a threshold where the emergent macroscopic behavior changes qualitatively. The hypothesis that the brain is operating in such a critical state is attractive because criticality is known to bring about optimal information processing and computational capabilities \cite{Beggs2003, Kinouchi2006, Legenstein2007, Larremore2011}. Recent experimental observations of patterns of neuronal activity exhibiting scale-free distributions, a typical hallmark of phase transitions, provided further evidence for this hypothesis. Bursts of neuronal activity were first shown in reduced preparations in rat brains to follow power-law probability distributions, termed neuronal avalanches \cite{Beggs2003, Beggs2004}. More recently, neuronal avalanches were also observed in invasive recordings from monkeys and cats, strongly suggesting that criticality is a generic property of cortical network activity \textit{in vivo} \cite{Gireesh2008, Petermann2009, Hahn2010}. 

Additional evidence for the existence of a critical state in human brain dynamics comes from a recent contribution by Kitzbichler et al. \cite{Kitzbichler2009}. Using magnetoencephalography (MEG) and functional magnetic resonance imaging (fMRI), the authors found power-law probability distributions of two measures of phase synchronization in brain networks. As confirmed by computational models, these distributions show power-law scaling specifically when those model systems are in a critical state resulting in strong evidence that human brain functional systems exist in an endogenous state of dynamical criticality at the transition between an ordered and a disordered phase. 

Theory predicts local events to percolate through the system in the form of avalanches of activity at the critical state \cite{Bak1995}. Such a critical state requires a homeostatic regulation of activity leading to a balance of excitation and inhibition in order to prevent states where events are either small and local or very large, engaging most of the network. 
A promising mechanism showing robust self-organized criticality (SOC) -- the ability of systems to self-tune their operating parameters to the critical state -- came from the discovery of network-based mechanisms, which were first reported in \cite{Christensen1998} and explained in detail in \cite{Bornholdt2000, Bornholdt2003}. These works showed that \textit{adaptive networks}, i.e., networks that combine topological evolution of the network with dynamics in the network nodes \cite{Gross2008}, can exhibit highly robust SOC based on simple local rules.
In computational models based on this adaptive interplay between network activity and topology, it could be shown that local, realistic synaptic mechanisms are sufficient to self-organize neuron networks to a critical state, providing a plausible explanation of how criticality in the brain can be achieved and sustained \cite{Levina2007, Meisel2009, Tetzlaff2010}. 
In summary, experimental findings in line with conceptual and computational models provide a plausible explanation of how criticality in the brain can be achieved and sustained.

A so far unresolved question concerns the medical relevance of critical brain activity. Diseases in the central nervous system are often associated with altered brain dynamics. 
It has been hypothesized that the dynamical properties characterizing a critical state may be seen as an important marker of brain well-being in both health and disease \cite{Expert2010}. 
Epilepsy is a malfunction of the brain associated with abnormal synchronized firing of neurons during a seizure \cite{Lehnertz2009}. 
The increased collective neuronal firing during attacks has been speculated to be linked to a pathological deviation away from a critical state \cite{Hsu2007}.
Evidence supporting this idea comes from recent \textit{in vitro} studies of animal brains.
There, application of receptor blockers could drive network dynamics away from its normal state where activity patterns of neuron dynamics deviated from a power-law \cite{Gireesh2008, Hsu2008}. 

Here, we confirm the previously observed critical dynamics with a complementary experimental methodology, providing additional evidence for the criticality hypothesis. Furthermore, we show that human brain networks \textit{in vivo} are not in a critical state during epileptic seizure attacks.
Deriving the distribution of phase-lock intervals (PLI) as an indicator of critical brain dynamics from electrocorticogram (ECoG) data, we find the distribution's scale invariance to be disturbed during seizures. Combined with results from a computational model exhibiting SOC these observations suggest the dynamical changes during seizure attacks to rely on a deviation from a critical state and hint to the failure of adaptive SOC as a cause for seizure generation. Based on the distribution of PLI we identify a measure which could prove useful for future seizure prediction algorithms.

\section{Materials and Methods}
\subsection*{Acquisition And Preprocessing Of Experimental Data}
Eight patients undergoing surgical treatment for intractable epilepsy participated in the study. Patients underwent a craniotomy for subdural placement of electrode grids and strips followed by continuous video and ECoG monitoring to localize epileptogenic zones. Solely clinical considerations determined the placement of electrodes and the duration of monitoring. All patients provided informed consent. ECoG signals were recorded by the clinical EEG system (epas 128, Natus Medical Incorporated) and bandpass filtered between 0.53 Hz and 70 Hz. Data were continuously sampled at a frequency of 200 Hz (patients 1-7) and 256 Hz (patient 8, \cite{Ihle2011}). The study protocols were approved by the Ethics Committee of the Technical University Dresden.

\subsection*{Estimation Of Phase Synchronization}
To derive a scale-dependent estimate of the phase difference between two time series, we follow the approach described in ref. \cite{Kitzbichler2009} using Hilbert 
transform derived pairs of wavelet coefficients \cite{Whitcher2005}. We define the instantaneous complex phase vector for two signals $F_i$ and $F_j$ as: 

\begin{equation}
\label{instphasevector}
C_{i,j}(t)=\frac{W_k(F_i)^{\dag}W_k(F_j)}{|W_k(F_i)||W_k(F_j)|},
\end{equation}

where $W_k$ denotes the $k$-th scale of a Hilbert wavelet transform and $^\dag$ its complex conjugate. A local mean phase difference in the frequency interval defined 
by the $k$-th wavelet scale is then given by 

\begin{equation}
\label{argc}
\Delta\phi_{i,j}(t)=Arg(\overline{C_{i,j}}),
\end{equation}

with 

\begin{equation}
\label{caverage}
\overline{C_{i,j}}(t)=\frac{\langle W_k(F_i)^{\dag}W_k(F_j)\rangle}{\sqrt{\langle|W_k(F_i)|^2\rangle \langle|W_k(F_j)|^2\rangle}}
\end{equation}

being a less noisy estimate of $C_{i,j}$ averaged over a brief period of time $\Delta t=2^k8$ \cite{Kitzbichler2009}. Intervals of phase-locking can then be identified as 
periods when $|\Delta\phi_{i,j}(t)|$ is smaller than some arbitrary threshold which we set to $\pi/4$ here. 

\subsection*{Estimation Of Deviation From A Power-law Distribution}
To quantify the deviation from a power-law we defined a measure $\Delta p$ similar to ref. \cite{Tetzlaff2010}. $\Delta p$ measures the difference between the cumulative density distribution of phase-lock intervals and a theoretical power-law distribution $p^{theo}$ obtained from a fit of the experimental data \cite{Clauset2009}.
$p^{theo}$ is calculated from the first time-interval (0-150 seconds) of a data set. For each time-interval of 150 seconds duration, $p^{theo}$ is then subtracted from the cumulative density distribution of PLI, $p^{orig}$, for each data point corresponding to a phase-lock interval $x$ and normalized by the number of data points $N$:

\begin{equation}
\label{deltap}
\Delta p=\frac{1}{N}\sum_x p_x^{orig}-p_x^{theo}.
\end{equation} 

Positive values of $\Delta p$ indicate a deviation with increased intervals of phase-locking, negative values indicate decreased phase-locking compared to the reference power-law distribution.

\subsection*{Computational Model}
An influential model explaining how dynamical systems can self-organize towards a critical state was introduced in ref. \cite{Bornholdt2000}. The mechanism is based on the
adaptive interplay between the dynamics of the nodes in the network (dynamics on the network) and the rewiring of the network's topology (dynamics of the network). More precisely, the 
topology of the network is changed according to the activity of the nodes in the network so that on average active nodes loose links and frozen nodes grow links. This local rewiring leads to a robust evolution towards a critical connectivity $K_c$ where the system is at a phase transition between order and disorder \cite{Bornholdt2000}. 

We first instantiated this model in a network of 200 randomly interconnected binary elements with states $\sigma_i=\pm1$ which are updated in parallel and scanned for local rewiring of connections. Under the algorithm described in ref. \cite{Bornholdt2000} the network's topology evolves towards a critical connectivity of $K_c\sim3.1$. Our objective was to get an estimate of phase-lock intervals between the activity of pairs of nodes for different average connectivities to provide a reference for comparable analysis neurophysiological time series. We therefore monitored states $\sigma_i(t)$ of 20 randomly chosen nodes in random networks instantiated at three different average connectivities: $K=2.75$, $K=3.1$ (critical) and $K=5.0$. Each network run was limited to 40000 time steps with no topological rewiring applied. From these data on the activity in the network cumulative distribution of PLI were calculated as described above. We found that the probability distribution of phase-lock intervals demonstrated power-scaling specifically when the system was at the critical connectivity $K=3.1$ whereas distributions at $K=2.75$/$K=5.0$ deviated from a power-law showing periods of increased/decreased phase-locking.

\section{Results}

We investigated data sets from ECoG acquired during presurgical monitoring of patients suffering from focal epilepsy. 
Data were continuously sampled at 200 Hz (patients 1-7) or 256 Hz (patient 8) with the number of channels ranging from 30 to 45 for different patients. The time series recorded from the anatomical site where the epileptic focus was assumed typically included one or more neurographically-identifiable seizure attacks. 

To test brain dynamics for signatures of criticality we analyzed ECoG activity in different time windows. The data sets were split in intervals of 150 seconds length (30000 sample steps at 200 Hz sampling, 38400 in the case of 256 Hz) with consecutive intervals overlapping by 100 seconds (20000 sample steps at 200 Hz, 25600 at 256 Hz). Following the approach in \cite{Kitzbichler2009}, we determined the distribution of phase-locking intervals (PLI) as an experimentally accessible indicator of critical brain activity. 
The length of time windows was chosen to be long enough to give a reliable estimate of the distribution of PLI on the one hand and allow observation of its evolution in time on the other hand.
For each of these sets, we calculated phase-lock intervals and determined their cumulative density distributions for scales 2, 3 and 4 corresponding to frequency intervals 50-25 Hz, 25-12.5 Hz and 12-6 Hz for patients 1-7 (P1-P7) and 64-32 Hz, 32-16 Hz, 16-8 Hz for patient 8.

The distributions for all scales closely follow a power-law probability distribution with $p(PLI) \sim PLI^{\alpha}$ during pre-ictal time intervals. Statistical tests based on the Kolmogorov-Smirnov statistic and likelihood ratios \cite{Clauset2009} showed that the hypothesis of a power-law PLI distribution could not be rejected for most pre-ictal data sets, furthermore a recent comprehensive analysis of various fitting functions applied to PLI distributions had revealed a power-law to be the most likely fit \cite{Kitzbichler2009}. 
The apparent robustness of the power-law against exact conditions (different anatomical regions with varying number of channels) strengthens the hypothesis of the relevance of a critical state in human brain dynamics.

\begin{figure}
\includegraphics[width=0.5\textwidth]{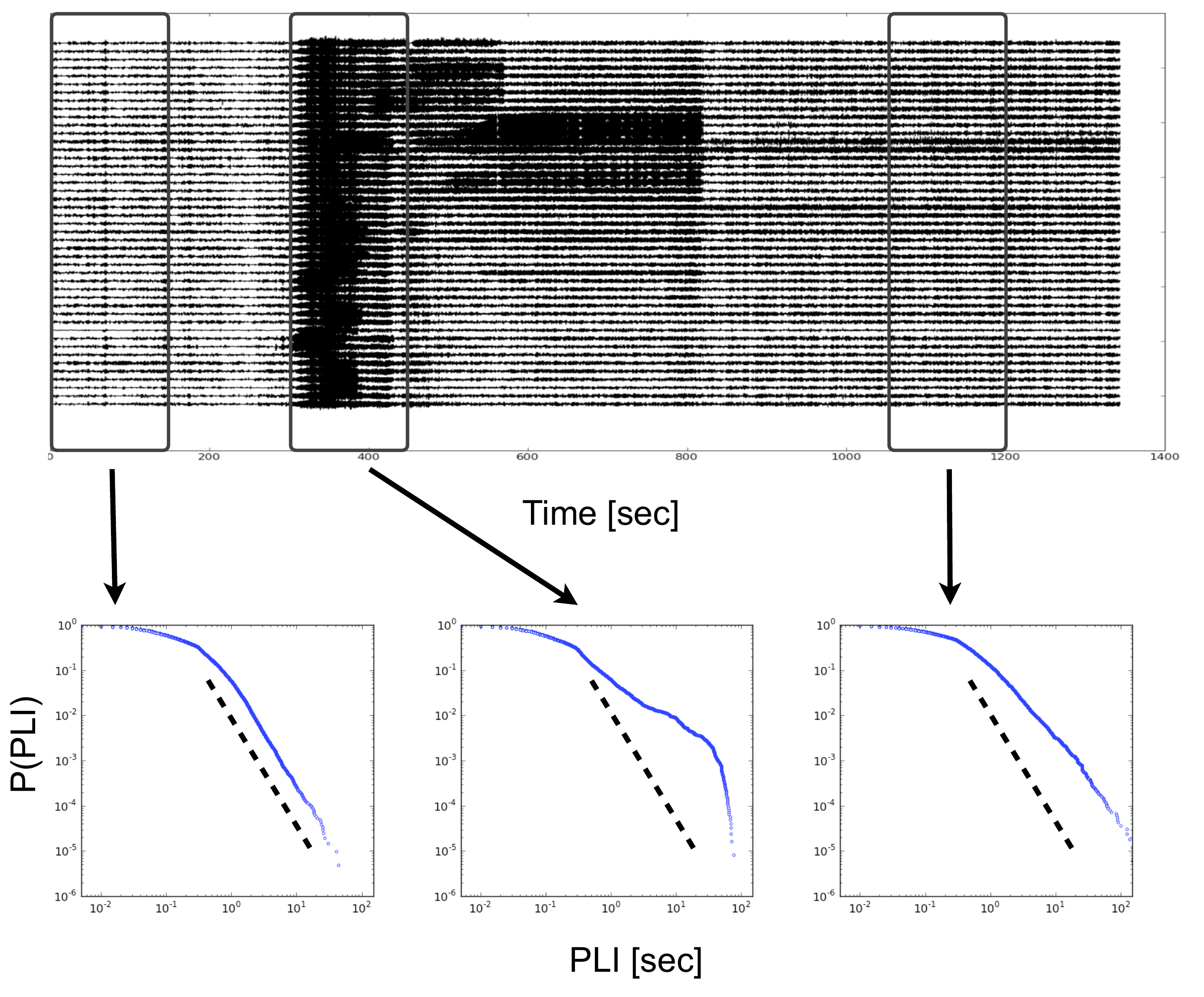} 
\caption{The distribution of phase-locking intervals deviates from a power-law during epileptic seizure attacks. Top: The electrocorticogram (ECoG) recording shows the onset of a focal epileptic seizure attack around 300 seconds time. Bottom: Cumulative distributions of phase-locking intervals (PLI) are obtained during three time intervals of 150 seconds: pre-ictal (left), ictal (middle) and post-ictal (right). Dashed lines indicate a power-law in the double logarithmic plot.
While the distribution appears to follow a power-law during the pre-ictal period, intervals of increased phase-locking disturb this characteristic distribution with the onset of seizure activity. Data shown are from patient 1 at scale 3, corresponding to the frequency band 25-12.5 Hz.}\label{fig1}
\end{figure}

\begin{figure}
\includegraphics[width=0.4\textwidth]{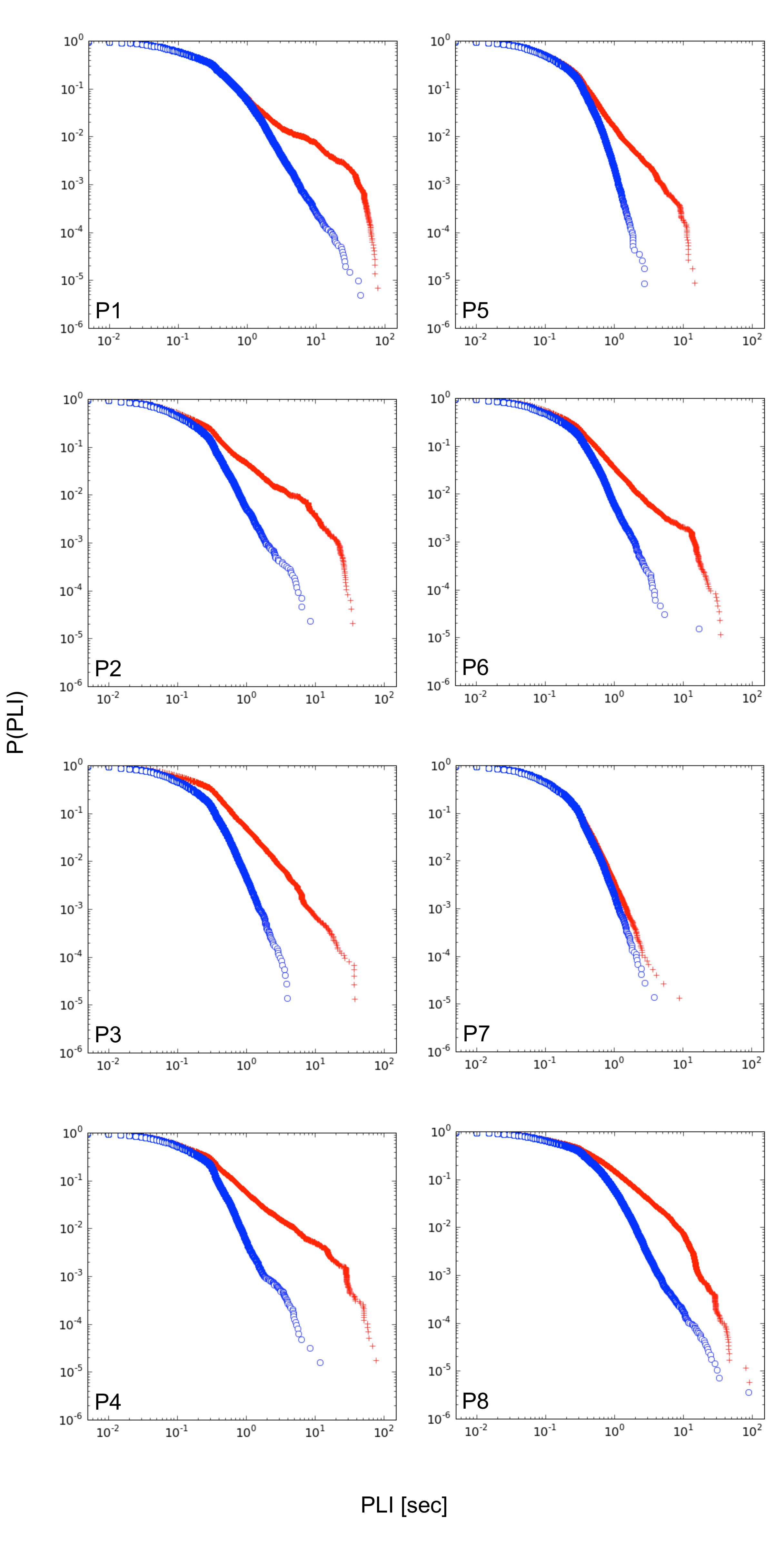} 
\caption{Comparison of PLI distributions derived from the first (pre-ictal) time interval (blue curve) and an interval during the seizure attack (red curve). Distributions from seizure intervals tend to exhibit longer periods of phase-locking resulting in a deviation from a power-law of the distribution's tail. Plots are shown for scale 3 corresponding to the frequency band 25-12.5 Hz for patients 1-7 (P1-P7) and 32-16 Hz for patient 8 (P8), respectively.}\label{fig2}
\end{figure}

While the PLI distribution followed a power-law in time intervals preceding the seizure onset, a deviation from power-law behavior was observed in intervals containing the seizure attack. 
Figure \ref{fig1} shows distributions of PLI derived from a pre-ictal, an ictal and a post-ictal time interval. The probability to find longer PLI tended to be increased during attacks thereby destroying the scale-free property of the original distribution. This qualitative change away from a power-law distribution could be observed in all 8 patients and across scales (Fig. \ref{fig2}).

A more quantitative estimate of the deviation from the pre-ictal state can be obtained by calculating $\Delta p$, a measure previously proposed to characterize the divergence from a critical state \cite{Tetzlaff2010}. 
This measure captures the deviation from a given empirical distribution from a power-law. The power-law fitted to the first (pre-ictal) interval was thereby taken as a reference and subtracted from the cumulative PLI distributions of subsequent time intervals. 

\begin{figure*}
\includegraphics[width=0.8\textwidth]{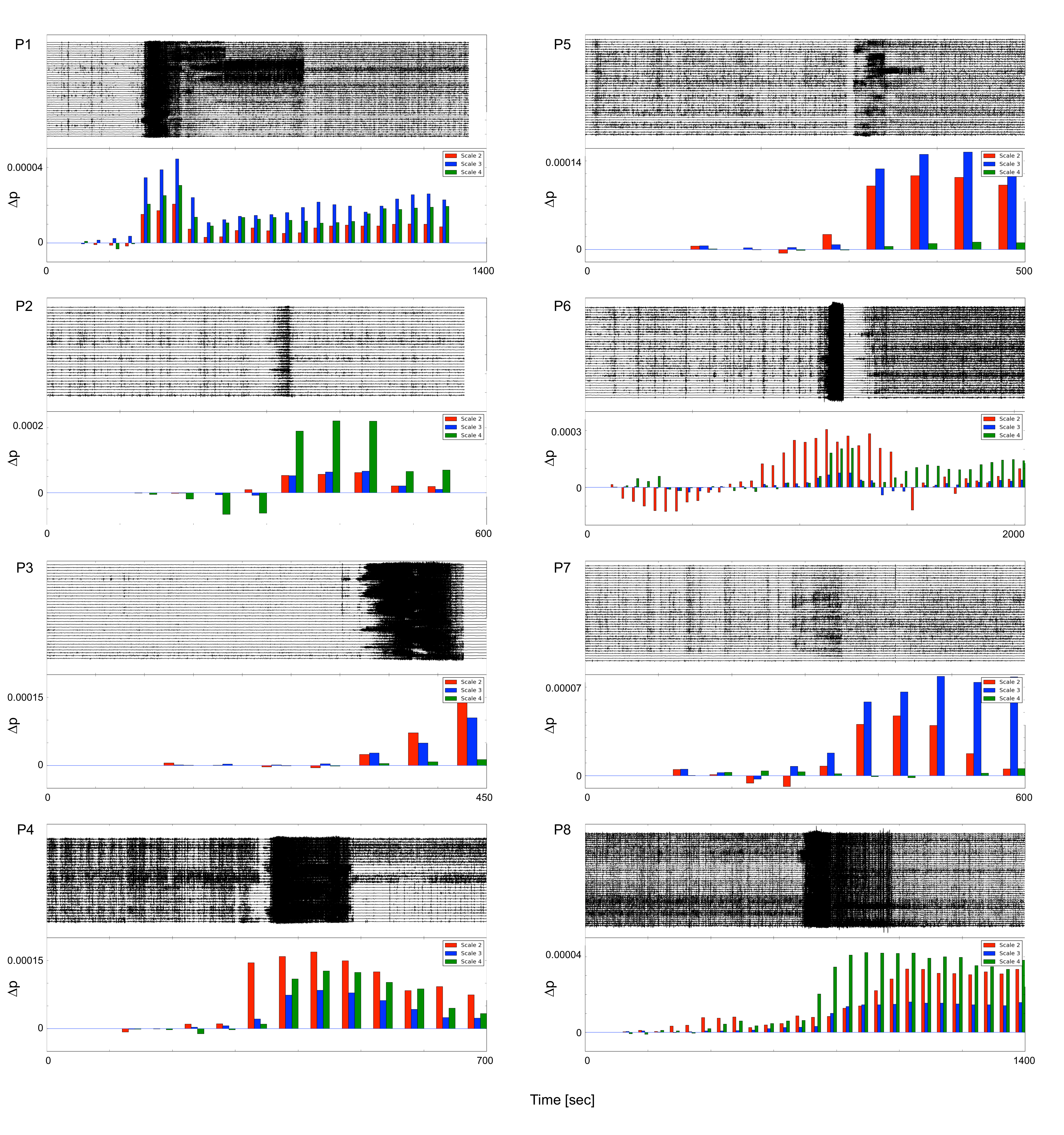} 
\caption{Development of the deviation from a power-law. ECoG recordings from 8 patients showing a focal seizure attack are shown along with $\Delta p$ values for consecutive time windows of 150 seconds duration overlapping by 100 seconds. The power-law fit of data in the first time window was taken as the reference to calculate $\Delta p$. Although different in extent, an increase of $\Delta p$ quantifying the deviation from the initial pre-ictal distribution can be observed during seizures for all patients and different scales.}\label{fig3}
\end{figure*}

During time intervals preceding the seizure $\Delta p$ stayed at low values indicating no significant deviation from a power-law. In time windows containing seizure activity, $\Delta p$ increased to positive values, which is in agreement with the qualitative assessment from visual inspection showing a divergence from the initial distribution. After seizure attacks, a slow decrease of $\Delta p$ could be observed suggestive of a relaxation process back toward a power-law distribution (Fig. \ref{fig3}).

\begin{figure}
\includegraphics[width=0.4\textwidth]{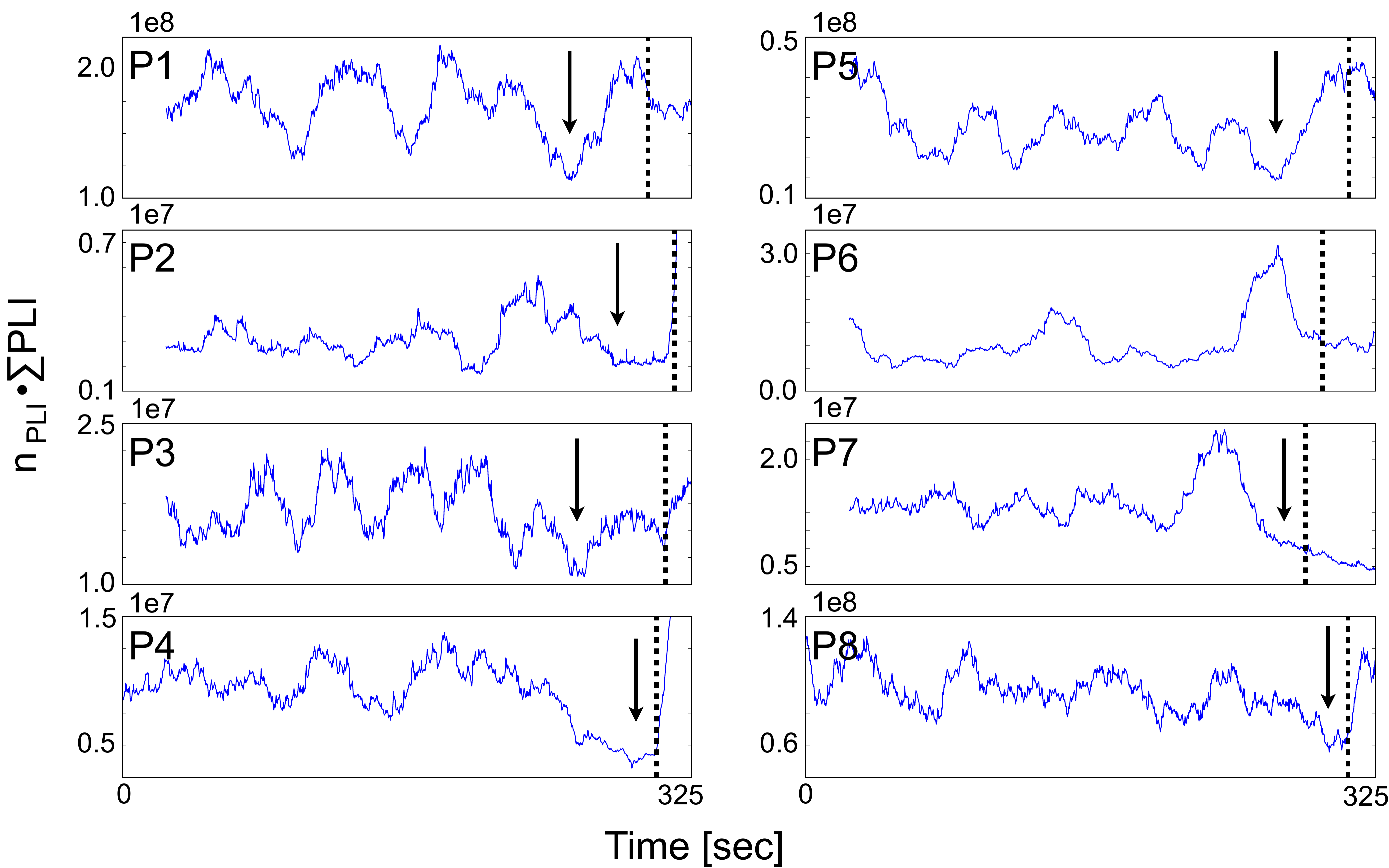} 
\caption{Evolution of the amount of phase-locking quantified by $n_{PLI}\cdot\sum{PLI}$, with $n_{PLI}$ being the number and $\sum{PLI}$ the sum of all phase-lock intervals. The measure varies during the time course and tends to decrease reaching a minimum (marked by black arrows) shortly before the beginning of the seizure. Seizure onset times are marked by vertical dashed lines. Phase-locking intervals were determined from overlapping time windows of 5000 sampling steps length for the frequency band defined by scale 4.}\label{fig5}
\end{figure}

Although $\Delta p$ values hinted that the power-law distribution was conserved during pre-ictal intervals, we observed variations in the amount of phase-locking as time approached the seizure onset. To investigate this with a better resolution in time we repeated the calculation of PLI for shorter time windows of 5000 sampling steps (corresponding to 25 seconds for patients 1-7 and 19.53 seconds for patient 8) every 50 sampling steps (corresponding to 0.25 or 0.19 seconds). While the power-law behavior of the PLI distribution appeared to be conserved until seizure onset, we found that the number $n_{PLI}$ and sum of all phase-lock intervals $\sum{PLI}$ varied during pre-ictal periods. 
Figure \ref{fig5} shows the product of the two, $n_{PLI}\cdot\sum{PLI}$, as a time series for each patient. 
Furthermore, this measure tended to decrease toward the time of seizure onset. For the scales 2-5 under investigation, the decrease was occurring most prominently in the frequency band of scale 4 (12-6 Hz P1-P7 and 16-8 Hz P8). 
While the amount of phase-locking varied and showed the tendency to decrease toward seizure onset times, the power-law tail of the PLI density distribution was conserved indicating to a robust underlying mechanism producing the power-law distribution. 

For obtaining further insights into the underlying dynamics of the power-law probability distribution of PLI and its absence during epileptic seizure attacks, we compared experimental results to a simple computational model exhibiting self-organized criticality. 
The network model introduced by Bornholdt and Rohlf \cite{Bornholdt2000} self-organizes toward a critical state and through its simplicity allows for an understanding of the underlying mechanism by which this self-tuning is accomplished. 
Specifically, the adaptive interplay of network dynamics and topology, a mechanism also at work in more elaborate models of SOC in neural networks \cite{Levina2007, Meisel2009, Tetzlaff2010}, robustly organizes systems parameters, in this case the average connectivity $K$, toward values $K_c$ where the network's state is at a phase transition between ordered and disordered dynamics. 

\begin{figure}
\includegraphics[width=0.4\textwidth]{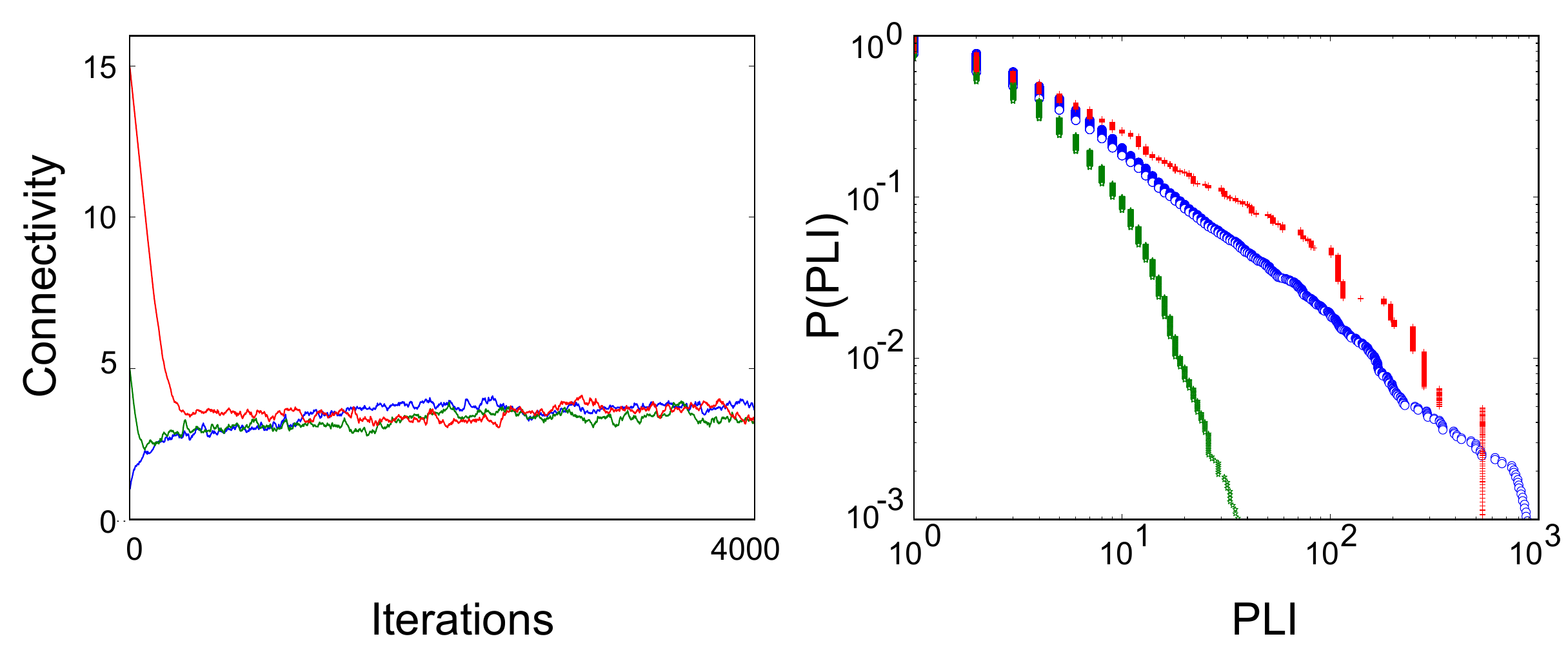} 
\caption{Distribution of PLI in a model exhibiting self-organized criticality. Through an adaptive interplay of network dynamics and topology, the Bornholdt model evolves toward a critical state with an average connectivity $K_c$. The left graph shows the evolution to $K_c\sim3.1$ for a network with 200 nodes for different initial connectivities. Right: cumulative distributions of PLI for scale 1 in random networks at average connectivities $K=5.0$ (green), $K=3.1$ (blue) and $K=2.75$ (red). The PLI distribution follows a power-law at the self-organized connectivity $K_c=3.1$.}\label{fig4}
\end{figure}

At the self-organized connectivity, statistical tests \cite{Clauset2009} revealed that the hypothesis of a power-law for the distribution of PLI can be accepted whereas it has to be rejected for distributions at connectivities below or above $K_c$ (Fig. \ref{fig4}). There, the distribution of PLI deviated from a power-law consistent with a state away from critical dynamics. 
The distribution at connectivities corresponding to the ordered phase of network dynamics is shifted towards larger PLI similar to the one observed during epileptic seizure attacks (bottom right in Fig. \ref{fig4}). 

The close agreement between patient and model data suggests that the deviation from a power-law observed during epileptic seizure attacks relies on a shift of dynamics toward an ordered phase.  
It further hints that it is the mechanism of adaptive SOC, the ability to tune system parameters to values where network dynamics is at a phase transition and PLI are distributed according to a power-law, that fails during epileptic seizure attacks in neuron networks in the brain. 

\section{Discussion}
The relevance of critical brain dynamics is currently a heavily debated topic. Indirect evidence for such a state comes from power-law distributed observables in neurophysiological data. Although various mechanisms can result in an event size distributions exhibiting power-laws \cite{Newman2005}, such distributions arise when a system is in a critical state \cite{Bak1995}. Recently, the power-law distribution of phase-lock intervals between pairs of neurophysiological time series was shown to be a specific hallmark of dynamic criticality in human brain dynamics \cite{Kitzbichler2009}. Using this indicator on ECoG data, a complementary experimental methodology to \cite{Kitzbichler2009}, we confirm the previously observed critical dynamics, providing additional evidence for the criticality hypothesis. Secondly, we show that the critical state is disturbed during epileptic seizure attacks. More precisely, the distribution of the PLI synchronization measure deviates from a power-law, characterizing the critical state of normal neuronal dynamics, during epileptic seizures, providing the first direct evidence of disturbed critical dynamics related to a pathology \textit{in vivo}.

Our findings support the notion of a physiological default state of balanced brain dynamics between regimes of exuberant and frozen activity. Physiological neuronal activity is characterized by intermittent periods of synchronization between different anatomical regions. In terms of dynamical system's theory, such a state corresponds a critical state at a phase transition between order and disorder. A deviation from this balanced state toward dynamics with pathologically increased times of synchronous activity as observed in epileptic patients leads to a deviance from the physiological critical state resulting in impaired functionality. 

Optimal functional capabilities of neuron networks have been related to a critical state before \cite{Beggs2003, Haldeman2005, Chialvo2010}.  
In this work we observed that power-law scaling of phase-lock intervals is conserved until the seizure onset at which the power-law character of the PLI distribution is lost relatively abruptly. This is remarkable since other measures characterizing the PLI distribution such as the sum and number of all PLI are varying and tending to decrease prior to the seizure.  
With power-law scaling as an indicator of criticality this suggests that a state of critical dynamics is maintained until the seizure onset and lost thereafter which manifests in impaired functionality and the start of clinical symptoms. 

In comparison to the relatively late loss of the characteristic power-law distribution, other precursors are much earlier indicative of a seizure onset. The identification of precursors of a seizure attack is a question of high clinical relevance \cite{Mormann2007}. Earlier studies reported a decrease of synchronization-based measures suggesting them as promising indicators of a pre-ictal state \cite{Mormann2003a, Mormann2003b, VanQuyen2005, Schelter2006}. 
We also observed a tendency of decreased phase-locking prior to the seizure attack. In a recent study using a measure based on phase-locking the authors found pre-ictal decreases of synchronization levels to be occurring mostly within the 4-15 Hz band \cite{VanQuyen2005} similar to our findings which were most evident for scale 4. Given the amount of information on phase-locking in a specific frequency band contained in our measure $n_{PLI}\cdot\sum{PLI}$ one can speculate that it should perform better in detecting pre-ictal changes of neural dynamics than other less specific synchronization measures. However, it should be noted that in order to assess the usefulness and performance a statistical validation taking into account the sensitivity and specificity of this measure is needed which is beyond the scope of this article.

A mechanism by which complex networks can self-organize toward a critical state is based on the adaptive interplay between the dynamics \emph{on} the network, i.e.\ neuronal activity, and the dynamics \emph{of} the network, i.e.\ the shaping of synaptic connections. Through this interaction system parameters can be locally tuned to a state of global criticality \cite{Bornholdt2000, Bornholdt2003}. While the simple model described in this work captures these essential ingredients allowing for an understanding of the underlying concept, more elaborate mechanism can be expected to be at work in real-world neuron networks \cite{Levina2007, Meisel2009, Tetzlaff2010, Millman2010}. It is conceivable that physiological neuron networks in the brain tune their parameters to more than one parameter to reach a state of criticality. Besides the average connectivity of the network $K$, the balance between excitation and inhibition, for example, has been shown to be an important parameter to sustain a homeostatically balanced critical state and prevent regimes of overly synchronized activity. 
It is conceivable that it is the robust mechanism of adaptive SOC maintaining neuron networks in the brain close to a critical state characterized by dynamics exhibiting power-law probability distributions even while network dynamics undergoes changes as captured by the variations of overall phase-locking, for example.  

The deviation from a power-law distribution of PLI reported here corresponds to a shift away from a balanced critical state and to our knowledge constitutes the first direct proof of impaired critical dynamics related to a pathology \textit{in vivo}. This observation is supported by experimental results from \textit{in vitro} studies. The application of receptor blockers in slice preparation of animal brains resulting in an excess of excitation in the network destroyed the power-law distributed avalanches of neuronal activity and led to increased avalanche sizes corresponding to a super-critical state \cite{Beggs2004, Gireesh2008}. Analogously, human tissue removed from epilepsy patients exhibited abnormally regulated avalanches with periods of hyperactivity \cite{Hobbs2010}. 

In summary, experimental results from \textit{in vitro} experiments \cite{Beggs2004, Gireesh2008} and \textit{in vivo} observations presented here combined with insights from computational models based on adaptive SOC \cite{Bornholdt2000, Levina2007, Meisel2009, Tetzlaff2010, Millman2010} suggest the failure of the adaptive interplay between neuron activity and network topology to lead to the deviation from a critical state with pathological, in the case of epilepsy overly synchronized, activity patterns. The sudden loss of the power-law distribution during the onset of an epileptic seizure attack hints to some threshold after which adaptive SOC is not able to maintain a critical state and fails. 
A deviation from the default critical state towards a dynamical regime with decreased phase-locking is also conceivable. For instance in neurodegenerative diseases with impaired neuronal connectivity, the deviation from a power-law of PLI could potentially be used to identify and characterize pathological conditions. 

\section*{Acknowledgments}
C.M. thanks M. Kirsch and E. Noback for their support in preparing data sets for analysis. We further thank M. Ihle and A. Schulze-Bonhage for supplying the recordings of patient 8.

\bibliography{my}

\end{document}